\documentstyle[preprint,eqsecnum,cite,epsf,aps]{revtex}
\begin{document}
\tightenlines

\title{\Large The Noncommutative  Supersymmetric Nonlinear Sigma  Model}

\author{H. O. Girotti$^{\,a}$, M. Gomes$^{\,b}$,
  V. O. Rivelles$^{\,b}$ and A. J. da Silva$^{\,b}$}  
\address{$^{a\,}$Instituto de F\'\i sica, Universidade Federal do Rio Grande
do Sul\\ Caixa Postal 15051, 91501-970 - Porto Alegre, RS, Brazil\\ E-mail:
hgirotti@if.ufrgs.br}
\address{$^{b\,}$Instituto de F\'\i sica, Universidade de S\~ao Paulo\\
 Caixa Postal 66318, 05315-970, S\~ao Paulo - SP, Brazil\\
E-mail: mgomes, rivelles, ajsilva@fma.if.usp.br}

\maketitle

\begin{abstract}
We show that the noncommutativity of  space--time destroys
the renormalizability of the $1/N$ expansion of the $O(N)$
Gross--Neveu model.  A similar statement holds for the noncommutative
nonlinear sigma model.  However, we show that, up to the subleading
order in $1/N$ expansion,  the noncommutative supersymmetric $O(N)$
nonlinear sigma model becomes renormalizable in $D=3$. We also show that
 dynamical
mass generation is restored and there is no catastrophic UV/IR mixing.
Unlike the commutative case, we find that the Lagrange multiplier
fields, which enforce the supersymmetric constraints, are also
renormalized. For $D=2$ the divergence of the four point function of
the basic scalar field, which in $D=3$ is absent, cannot be eliminated
by means of a counterterm having the structure of a Moyal product.

\end{abstract}

\newpage
\section{INTRODUCTION}
Recently, a great amount of interest has been devoted to the subject
of noncommutative field theory. This interest has various origins as
quantum gravity, string theory or just the investigations of the
conceptual basis of field theory. It was found that, contrary to
initial expectations, noncommutative models based on Moyal products
are in general plagued with inconsistencies like the breaking of
unitarity and causality \cite{Toumbas} and the mixing of
infrared/ultraviolet
divergences\cite{Minwalla,Matusis,Hayakawa,Arefeva2,Arefeva,Martin,Grosse,Armoni,Bonora,Sheikh,Chepelev,Gracia}. Whenever
possible, one could evade the problems of unitarity and causality by
restricting the noncommutativity to the space coordinates only.

Noncommutative field theories are nonlocal field models where the
nonlocality is a well defined consequence of the noncommutativity.
For a given model it could happen that the ultraviolet divergences do
not preserve the nonlocal Moyal structure of the bare vertices and the model
turns out to be nonrenormalizable though it is renormalizable in
ordinary space.  After verifying renormalizability, one should yet
determine up to what extent a noncommutative model retains the main
features of its commutative counterpart.  The answer to both questions
depends on the details of the underlying interactions.  For instance,
it was recently  shown that the $O(N)$ symmetry of the
noncommutative linear sigma model can not be
spontaneously broken for $N>2$, while, at least up to one loop, the
same does not apply for   the noncommutative $U(N)$ linear
sigma model when the ordering of the quartic  interaction is gauge
invariant\cite{Campbell}.

Most of the investigations on noncommutative
field theories  have been restricted to the first terms of the
perturbative series. Extension to higher orders are hampered
by technical difficulties  and  the UV/IR mixing. This last
feature, namely, the entanglement of scales which
mixes the ultraviolet and infrared behaviors leads to the
breakdown of the perturbative scheme in many of the standard
renormalizable theories. Fortunately, supersymmetric models \cite{Chu} appear to
be free from this drawback as it was shown  for the
Wess--Zumino model \cite{Girotti}. 

The question now is to decide whether the supersymmetric extension of
a noncommutative field theory not only preserves the renormalizability
but also other essential properties that its commutative
counterpart may eventually possess. 

In the present work we will use the $D$ dimensional ($2\leq D <4 $)
$O(N)$ Gross-Neveu (GN) model to illustrate these points. Although
perturbatively renormalizable only for $D=2$, its commutative version
is $1/N$ expandable and exhibits dynamical mass generation for both
$D=2$ and $D=3$ \cite{Neveu,Gross,Rosenstein,Gomes}. As we shall see,
noncommutativity breaks both  these aspects. Nevertheless, as in 
the Wess--Zumino model \cite{Girotti}, supersymmetry allows
to recover renormalizability and dynamical mass generation.

The supersymmetric partner of the GN model is the nonlinear sigma
model whose noncommutative extension is afflicted with UV/IR
mixing. As will be seen, supersymmetry also corrects this problem. It
should be stressed, however, that, due to the nonlocal character of
the Moyal product, the renormalization program for the noncommutative
supersymmetric nonlinear sigma model presents new aspects, which are
not shared by its commutative counterpart. In particular, as we shall
prove, multiplicative renormalizations for the auxiliary fields
become mandatory to achieve finite radiative corrections.

The paper is organized as follows. In section II we start by stressing
some basic aspects of the $1/N$ expansion of the commutative GN model
which will prove crucial for our discussion. Afterwards, we introduce
its noncommutative version and explicitly demonstrate the breakdown of
renormalizability. In this section we also discuss the UV/IR mixing in
the $1/N$ expansion of the nonlinear sigma model.  In section III we
pinpoint the tools needed to construct the $1/N$ expansion of the
supersymmetric nonlinear sigma model, which contains both the
nonlinear sigma model and the GN model
\cite{Alvarez,Witten,Arefeva1,Dadda,Davis,Koures}.  The noncommutative
supersymmetric nonlinear sigma model is studied in section IV with
special emphasis on the mechanism of mass generation. We also include
in this section the computation of the leading corrections to the
propagators of the basic fields. In section V we complete the
renormalization program in $2+1$ dimensions, up to the subleading
order in $1/N$. Some remarks on the
renormalization of the $1+1$ dimensional model and the conclusions are
presented in section VI.

\section{The Gross-Neveu and the non linear sigma models}
The commutative  $O(N)$ GN model is specified by the Lagrangian density

\begin{equation}
{\cal L}=\frac{i}2\overline \psi \not \!\partial \psi +\frac {g}{4N}(\overline \psi
\psi)(\overline \psi\psi).\label{1}
\end{equation}

\noindent
where $\psi_i,\quad i=1,\ldots N$ are two-component Majorana
spinors. Only in two dimensions this model is perturbatively
renormalizable. Although perturbatively nonrenormalizable in 2+1
dimensions, the model is $1/N$ expandable and presents some interesting
aspects such as dynamical mass generation \cite{Gross,Rosenstein,Gomes}.  From now on our discussion
will be restricted to $2\leq D < 4$.

To implement the $1/N$
expansion one introduces an auxiliary field $\sigma$ 
which enables  to write the Lagrangian of the theory as

\begin{equation}
{\cal L} =\frac{i}2\overline \psi \not \!\partial \psi -\frac{\sigma}2 
(\overline \psi
\psi)- \frac{N}{4g}\sigma^2.\label{2}
\end{equation} 

\noindent
At the quantum level one should integrate over $\sigma$ which can
develop a nonvanishing vacuum expectation value (VEV). We replace
$\sigma$ by $\sigma+M$ where $M$ is the VEV of the
original $\sigma$. The new Lagrangian is

\begin{equation}
{\cal L} =\frac{i}2\overline \psi \not \!\partial \psi - \frac{M}2\overline 
\psi
\psi-\frac{\sigma}2 (\overline \psi
\psi)- \frac{N}{4g}\sigma^2-\frac{N}{2g} M \sigma.\label{3}
\end{equation} 

\noindent
Since, by construction, the new field $\sigma$  has zero VEV,
the gap equation

\begin{equation}
 \frac{M}{2g}-i\int \frac{d^Dk}{(2\pi)^D}\frac{M}{k^{2}-M^2}=0\label{4}
\end{equation}

\noindent
must be obeyed. After a Wick rotation one has

\begin{equation}
 \frac{M}{2g}-\int \frac{d^Dk}{(2\pi)^D}\frac{M}{k^{2}_{E}+M^2}=0.\label{5}
\end{equation}

\noindent
The ultraviolet divergence in the above integral may be eliminated by
means of a coupling constant renormalization. Indeed, by defining the renormalized
coupling constant through

\begin{equation}
\frac{1}{g}= \frac{1}{g_R} + 2 \int \frac{d^Dk}{(2\pi)^D}\frac{1}{k^{2}_{E}
+\mu^2},\label{6}
\end{equation}

\noindent
the divergence is canceled.
In 2+1 dimension one finds 
\begin{equation}
\frac{1}{g_R}= \frac{\mu-|M|}{2\pi}\label{7}
\end{equation}

\noindent
and therefore only for $-\frac{1}{g_R}+\frac{\mu}{2\pi}>0$ it is possible to
have $M\not =0$. Otherwise $M$ is necessarily zero. In two dimensions
no such restriction exists. Whatever the case, we shall only focus on the
massive phase.

We shall next compute the propagator for the auxiliary field.
It is given by $-1/F(p)$, where

\begin{eqnarray}
F(p)&=& -\frac {iN}{2g}-  N \int \frac{d^Dk}{(2\pi)^D} \frac{k\cdot (k+p) + 
M^2}{(k^2-M^2)[(k+p)^2-M^2]}\label{8}
\end{eqnarray}

\noindent
is the two point function of the sigma field.
This last integral is divergent but after taking into account (\ref{4})
it  becomes

\begin{eqnarray}
F(p)&= &\frac{(p^2-4M^2)N}2\int\frac{d^Dk}{(2\pi)^D} \frac{1}{(k^2-M^2)[(k+p)^2-M^2]},
\label{10}
\end{eqnarray}

\noindent
which is finite for $D<4$. We see that the above cancellation of divergences
results from a fine tuning between the divergence in the integral
in  (\ref{8}) and the one in the gap equation.

Let us turn our attention to the noncommutative version of the model. By
introducing Moyal products we arrive at

\begin{equation}
S_{GN}=\int d^Dx \left [\frac{i}2\overline \psi \not \!\partial \psi - \frac{M}2\overline \psi
\psi-\frac12\sigma \star(\overline \psi\star
\psi)- \frac{N}{4g}\sigma^2- \frac{N}{2g}M\sigma\right ].\label{11}
\end{equation} 

\noindent
As can be checked,  the path integration on $\sigma$ leads to a noncommutative
version of (\ref{1}) in which the four--fermion interaction is
$\overline \psi_i\star\psi_i\star\overline \psi_j\star\psi_j$. A more
general $O(N)$ noncommutative four--fermion interaction may involve the
term $\overline \psi_i\star\overline \psi_j\star\psi_i\star\psi_j$. However
this last combination does not have a simple $1/N$ expansion and it will not be considered in this work.

Since the Moyal product does not affect the quadratic part, the
propagator for $\psi$ is as before. On the other hand, in momentum
space the trilinear vertex has to be multiplied by $\cos(p_1\wedge
p_2)$ where $p_1$ and $p_2$ are the momenta through the fermion lines
\footnote{
Here we have introduced the notation $a\wedge b= 1/2 a^\mu b^\nu
\Theta_{\mu\nu}$, where $\Theta_{\mu\nu}$ is the anti-symmetric
constant matrix characterizing the noncommutativity of the underlying
space.}.  Then, the
gap equation is not modified whereas, for the proper two point function of 
the $\sigma$ field, one obtains

\begin{equation}
{\cal F}(p)= -\frac {iN}{2g}- N \int \frac{d^Dk}{(2\pi)^D} \cos^2(k\wedge p)
\frac{k\cdot (k+p) + 
M^2}{(k^2-M^2)[(k+p)^2-M^2]}\label{12}
\end{equation}

\noindent
and we see that the divergent parts do not match anymore. The
model is no longer  renormalizable. A similar conclusion was arrived in
\cite{Semenoff}.

An analogous situation arises in connection with the Goldstone theorem
within the context of  the  $O(N)$ linear sigma model.  Indeed, the pion
propagator counterterm, which is determined by the vanishing of the
VEV of the sigma field, is not modified by the
noncommutativity. However, for $N>2$, the graphs contributing to the one loop
corrections of the pion propagator are altered thus destroying the
renormalizability of the model \cite{Campbell}.

As a side observation, notice that, up to the leading order of $1/N$, no
inconsistency would arise had we employed Dirac instead of Majorana spinors.
This is so because the oscillating exponentials, characteristics of the
noncommutativity, cancel in the leading order contributions
to the self--energy parts. The same does not hold at higher
orders of $1/N$.

\begin{figure}
\centerline{\epsfbox{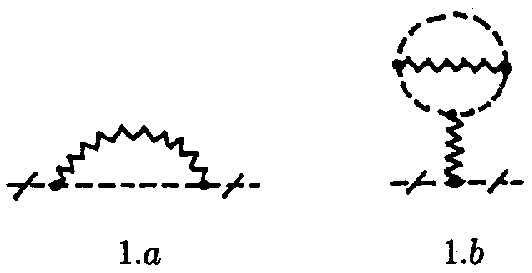}}
\caption{{Order $1/N$ contributions to the self--energy of the $\varphi$ 
field. Dashed and wavy lines\\
\hspace*{2.0cm}  represent, respectively, the propagators of $\varphi$ and $\lambda$ fields.}}\label{Fig1}
\end{figure}

We would like to remark that the noncommutative version of the
nonlinear sigma model, which happens to be the supersymmetric partner
of the GN model, also presents inconsistencies. The model is
specified by the Lagrangian

\begin{equation}\label{121}
{\cal L} =-\frac12\varphi(\partial^2+ M^2)\varphi + \frac12 \lambda\star
\varphi\star\varphi - \frac{N}{2g}\lambda,
\end{equation}

\noindent
where $\varphi_i$, $i=1,\ldots , N$ are real scalar fields,
 $M$ is, as before, the generated mass and $\lambda$ is an auxiliary
field which implements the nonlinear constraint $\varphi\star\varphi=N/g$.
The leading correction to the self-energy of the $\varphi$ field, shown in 
Fig. \ref{Fig1}$a$, is

\begin{equation}\label{122}
-i\int \frac{d^2 k}{(2\pi)^2} \frac{\cos^2 (k\wedge p)}{(k+p)^2-M^2}\Delta_\lambda(k),
\end{equation}

\noindent
where $\Delta_\lambda$ is the propagator for the $\lambda$ field. This
expression may be decomposed into a sum of a planar (quadratically divergent)
piece and a nonplanar one. The nonplanar
contribution is ultraviolet finite but diverges quadratically for
small momenta (UV/IR mixing \cite{Minwalla}) thus destroying the $1/N$
expansion of the model. Moreover, an attempt to solve this difficulty
by generalizing the definition of 1PI diagram \cite{Arefeva1} 
amounts to add the contribution from graph in Fig. \ref{Fig1}$b$ which is  given by

\begin{equation}\label{123}
-i \Delta_\lambda(0) \int  \frac{d^2 k}{(2\pi)^2} \frac{d^2 q}{(2\pi)^2}\frac{\cos^2 (k\wedge q)}{(k+q)^2-M^2}\frac{1}{(q^2-M^2)^2} \Delta_\lambda(k).
\end{equation}

\noindent
Like in the commutative case, the $\lambda$ field propagator
$\Delta_\lambda(k)$ behaves as $k^2$ when $k\rightarrow \infty$. Then,
for small $p$ the integral (\ref{122}) gives origin to a quadratic
infrared divergence which cannot be compensated by (\ref{123}). This
is the UV/IR mixing which, as already mentioned, breaks the $1/N$ expansion.

Before demonstrating that the noncommutative supersymmetric nonlinear
sigma model remains renormalizable, we shall recall some aspects of
its commutative counterpart
\cite{Alvarez,Witten,Arefeva1,Dadda,Davis,Koures}.

\section{The Supersymmetric Nonlinear sigma Model}

The commutative supersymmetric nonlinear sigma model is described by the Lagrangian density

\begin{equation}
{\cal L} =- \frac12\varphi \partial^2 \varphi + \frac{i}{2} \overline \psi
\not \! \partial \psi+ \frac12 F^2+ \sigma \varphi_j F_j + \frac12\lambda
\varphi^2 - \frac12\sigma \overline \psi \psi - \overline \xi \psi_j \varphi_j- \frac{N}{2g}\lambda,\label{13}
\end{equation}

\noindent 
where $\varphi_i,\psi_i$ and $F_i$, $i=1,\ldots, N$, are,
respectively, real scalar, two-component Majorana spinor and
auxiliary scalar fields. Furthermore, $\sigma,\lambda$ and $\xi$ are
Lagrange multipliers which implement the supersymmetric constraints

\begin{eqnarray}
\varphi_i\varphi_i &=& \frac{N}g,\\
\varphi_i \psi_i &=& 0,\\
\varphi_i F_i &=& \frac12 \bar \psi\psi.\label{14}
\end{eqnarray}

\noindent
A more symmetric form for (\ref{13}) is obtained after the linear
change of variables $\lambda\rightarrow \lambda + 2 M \sigma$,
$F\rightarrow F-M\varphi$ where $M=<\sigma>$. Afterwards, one performs
the shifts $\sigma\rightarrow \sigma +M$ and $\lambda\rightarrow
\lambda + \lambda_0$, where $\lambda_0=<\lambda>$, yielding

\begin{eqnarray}
{\cal L} &=&- \frac12\varphi( \partial^2+M^2) \varphi + \frac{1}{2} 
\overline \psi(i\not \! \partial - M)\psi+ \frac12 F^2+
 M^2 \varphi^2+ \frac12\lambda_0
\varphi^2   \nonumber\\ 
&\phantom a& + \frac12\lambda \varphi^2 +\sigma  \varphi_j F_j
 - \frac12\sigma \overline \psi \psi  - \overline \xi \psi_j \varphi_j- \frac{N}{2g}\lambda -\frac{N}{g}M\sigma.\label{15}
\end{eqnarray}

\noindent
Supersymmetry demands $\lambda_0=-2M^2$ and the gap equations
arising from  $<\lambda>=<\sigma>=0$,  for $M\not=0$, are found to
imply

\begin{equation}
\int \frac{d^D k}{(2\pi)^D} \frac{i}{k^2-M^2}= \frac{1}{g},\label{16}
\end{equation}

\noindent
so that as for the GN model an infinite coupling constant
renormalization is required. We must again investigate up to what
extent such normalization affects the computation of the propagator
for the sigma field. To that end we list the propagators for the basic
fields $\varphi$, $\psi$ and for the auxiliary field $F$. They are

\begin{mathletters}
\begin{eqnarray}
&& \Delta_{\varphi_i\varphi_j}(p)=\Delta_{ij}=\frac{i\delta_{ij}}{p^2-M^2},\\
&& S_F (p)= \frac{i\delta_{ij}}{\not \! p -M},\\ 
&& \Delta_{F_iF_j}= i\delta_{ij}.
\end{eqnarray}
\end{mathletters}

\noindent
Unlike  the GN model, the finiteness of the $\sigma$ field propagator
does not depend on the renormalization of the coupling constant $g$.
In fact, the proper part $F_\sigma$ of the two point function of the
sigma field receives contributions from the second and third
terms in the second line of (\ref{15}).  One finds that

\begin{eqnarray}
F_\sigma(p)&=& N \int \frac{d^Dk}{(2\pi)^D}\frac{1}{k^2-M^2}-
 N \int \frac{d^Dk}{(2\pi)^D}\frac{k\cdot (k+p)+ M^2}{[(p+k)^2-M^2][k^2-M^2]},\nonumber\\
&&\label{17}
\end{eqnarray}

\noindent
where the first term arises from the second order contribution of
the $\sigma \varphi F$ vertex and  the last one  originates from the second
order contribution of the $\sigma \overline \psi \psi$ vertex.
After a straightforward algebra one arrives at

\begin{equation}
F_\sigma(p)=\frac{(p^2-4M^2)N}2\int\frac{d^Dk}{(2\pi)^D}
 \frac{1}{(k^2-M^2)[(k+p)^2-M^2]}\,\,\label{18},
\end{equation}

\noindent
which is identical to (\ref{10}). We strongly remark that, in 
contradistinction to (\ref{10}), the finiteness 
here is not a consequence of a gap equation.

\section{ The Noncommutative Supersymmetric Nonlinear sigma  Model: Renormalization of the
two point functions}

The noncommutative supersymmetric nonlinear sigma model is specified by

\begin{eqnarray}
S&=&\int\left \{
- \frac12\varphi (\partial^2+M^2) \varphi + \frac{1}{2} \overline \psi
(i\not \! \partial -M)\psi+ \frac12 F^2+\frac{\lambda}{2}\star\varphi
\star\varphi  
\right.\nonumber \\
& \phantom a &\left. - \frac12 F_j\star (\sigma\star\varphi_j +\varphi_j\star\sigma)-\frac12 \sigma\star\overline\psi\star\psi 
-\frac12 (\bar\xi\star\psi\star\varphi +\bar\xi\star\varphi\star\psi)\right.
\nonumber\\
&\phantom a &\left.-\frac{N}{2g}\lambda - \frac{N M\sigma}{g}\right \}d^D x.
\label{19}
\end{eqnarray}

\noindent
 It should be noticed that the symmetrized forms used above are the only
noncommutative 
supersymmetric extensions for the  terms $ \sigma \varphi F$
and $\overline \xi\psi\varphi$. It must also be emphasized that
the noncommutative supersymmetry transformations are identical to the
commutative ones since they are linear in the fields and no Moyal
products are, therefore, involved.

Since the quadratic part of the action was not modified, the free
propagators of the basic fields $\varphi$, $\psi$ and $F$ remain unaltered. As for the
vertices, they  acquire cosine factors as follows 

\begin{mathletters}
\begin{eqnarray}\label{20}
\lambda \varphi^2 \quad {\mbox {vertex:}}&& \quad \frac{i}2 
\cos(p_1\wedge p_2), \\
\sigma \varphi F \quad  {\mbox {vertex:}} && \quad - i \cos (p_1\wedge p_2),\\
\overline \psi \psi \sigma \quad {\mbox {vertex:}} &&\quad -\frac{i}2 
\cos (p_1\wedge p_2),\\ 
\overline \xi \psi \varphi \quad {\mbox {vertex:}} &&\quad - i 
\cos (p_1\wedge p_2).
\end{eqnarray}
\end{mathletters}

\noindent
Using the above rules one can compute the leading order propagators
for the Lagrange multiplier fields. For the proper part of the
$\sigma$ field, $ {\cal F}_\sigma$, one merely obtains the expression
(\ref{17}) modified by the presence of the factor $\cos^2(k\wedge p)$
in each integral. Hence,

\begin{equation}
{\cal F}_\sigma(p)=\frac{(p^2-4M^2)N}2\int\frac{d^Dk}{(2\pi)^D}
 \frac{\cos^2(k\wedge p)}{(k^2-M^2)[(k+p)^2-M^2]}\label{21},
\end{equation}

\noindent
which is well behaved in both  the infrared and ultraviolet regions. 
The propagator for the $\sigma$ field is, of course,  $\Delta_\sigma= -1/{\cal F}_\sigma$.

The expressions for the $\lambda$ and $\xi$ propagators
are given by $\Delta_\lambda=-1/{\cal F}_\lambda$ and $S_\xi=-1/{\cal F}_\xi$, respectively, where 

\begin{equation} 
{\cal F}_\lambda(p)= \frac{N}{2}\int \frac{d^D k}{(2\pi)^D} \cos^2(k\wedge p)\frac{1}{[(k+p)^2-M^2] [k^2-M^2]}\label{22}
\end{equation}

\noindent
and

\begin{eqnarray}
&&{\cal F}_\xi(p)={N}\int \frac{d^D k}{(2\pi)^D}{\cos^2(k\wedge p)}
\frac{-\not \! k + M}
{[(k+p)^2-M^2][k^2-M^2]}\nonumber \\ 
&&={N} \frac{(\not \! p + 2 M)}2\int \frac{d^D k}{(2\pi)^D}
{\cos^2(k\wedge p)} \frac{1}{[(k+p)^2-M^2][k^2-M^2]},\label{23}
\end{eqnarray}

\noindent
which are ultraviolet finite and without singularities for small $p$.
We observe that all the Lagrange multiplier field propagators 
are of order $1/N$.

\begin{figure}
\centerline{\epsfbox{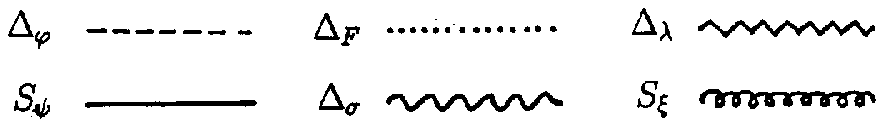}}
\caption{{Graphical representation for the propagators.}}
\label{Fig2}
\end{figure}

Our graphical notation for  the propagators is
presented in Fig. \ref{Fig2}. Observe that, being a constant, the  $F$ propagator  
could be omitted altogether, but one is not to
forget the cosine factors in the original graph.

Due to the oscillating nature of the cosines some of the
integrals constructed with the above rules will be finite but in general
divergences will survive. The degree of superficial divergence
for a generic 1PI graph $\gamma$ is

\begin{equation}
d(\gamma)= D - \frac{(D-1)}2N_\psi- \frac{(D-2)}2 N_\varphi-\frac{D}2 N_F- N_\sigma- \frac32 N_\xi-
2 N_\lambda,\label{24}
\end{equation}

\noindent
where $N_{\cal O}$ is the number of external lines associated to the 
field ${\cal O}$. Potentially dangerous diagrams are those contributing
to the self--energies of the $\varphi$ and $\psi$ fields since, in principle,
they are quadratic and linearly divergent, respectively.

\begin{figure}
\centerline{\epsfbox{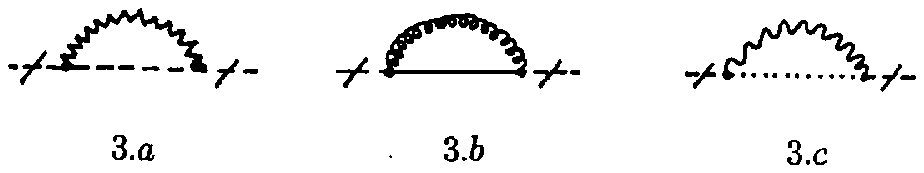}}
\caption{{$1/N$ corrections to the $\varphi$ field self--energy.}}
\label{Fig3}
\end{figure}

In lowest order there are three graphs contributing to the $\varphi$
field self--energy which are shown in Fig.\ref{Fig3}. In a self explanatory notation, 
the analytic expressions associated with them are

\begin{equation}
\Sigma_{\varphi}^a=-i\int\frac{d^Dk}{(2\pi)^D} \frac{\cos^2(k\wedge p)}{(k+p)^2-M^2}
\Delta_{\lambda}(k),\label{25}
\end{equation}

\begin{equation}
\Sigma_{\varphi}^b=i\int\frac{d^Dk}{(2\pi)^D}{\rm Tr}\left \{ \frac{\cos^2(k\wedge p)}{(\not\!\! k+\not\!\! p )-M}\,
\frac{1}{\not\!
\! k + 2M}\right \}\Delta_{\lambda}(k),\label{26}
\end{equation}

\noindent
and

\begin{equation}
\Sigma_{\varphi}^c=-i\int\frac{d^Dk}{(2\pi)^D} \frac{\cos^2(k\wedge p)}{k^2 - 4 M^2}\Delta_{\lambda}(k).\label{27}
\end{equation}

\noindent 
By adding these three expressions we get

\begin{equation}
\Sigma_{\varphi}=-i(p^2 -M^2) \int \frac{d^Dk}{(2\pi)^D}\frac{\cos^2(k\wedge p)}{[(k+p)^2-M^2][k^2 - 4 M^2]}\Delta_{\lambda}(k)\label{28}.
\end{equation}

\begin{figure}
\centerline{\epsfbox{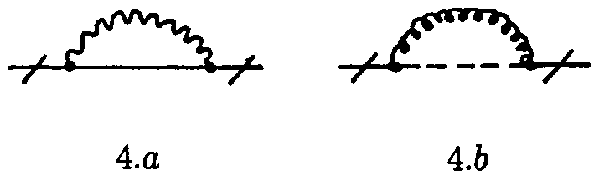}}
\caption{{$1/N$ corrections to the $\psi$ field self--energy.}}
\label{Fig4}
\end{figure}

\noindent
Individually each of the graphs  in Fig. \ref{Fig3} is quadratically divergent but their
sum diverges only logarithmically. This divergence can be eliminated
by a wave function renormalization of the $\varphi$ field. Notice
that, in spite of the presence of ${\cos}^2(k\wedge p)$, the divergent
part of (\ref{28}) coincides with the corresponding one in the
commutative case.  Similar result holds for the self--energy of the
$\psi$ field. Indeed, from Fig. \ref{Fig4} we obtain

\begin{equation}
\Sigma_{\psi}^a=-i\int \frac{d^Dk}{(2\pi)^D} 
\frac{(\not\!\! k+\not\!\! p+M)
\cos^2(k\wedge p)}{[(k+p)^2-M^2][k^2 - 4 M^2]}\Delta_\lambda(k)\label{29}
\end{equation}

\noindent
and

\begin{equation}
\Sigma_{\psi}^b=i \int \frac{d^Dk}{(2\pi)^D}
\frac{(\not\!\! k+2 M) 
\cos^2(k\wedge p)}{[(k+p)^2-M^2][k^2 - 4 M^2]}\Delta_\lambda(k),\label{30}
\end{equation}

\noindent
so that

\begin{equation}
\Sigma_\psi= -i(\not\!\! p-M) \int \frac{d^Dk}{(2\pi)^D}
\frac{\cos^2(k\wedge p)}{[(k+p)^2-M^2][k^2 - 4 M^2]}\Delta_\lambda(k).
\label{31}
\end{equation}

\noindent
We see that the leading divergence is again canceled and just remains
a logarithmic one which may be absorbed by a wave function
renormalization of the $\psi$ field. A similar analysis of the
logarithimic divergence present in the $F$ field propagator reveals
that it can be also removed by a wave function renormalization of the $F$
field.

We stress that the renormalization
factors for the fields  $\varphi$, $\psi$ and  $F$ are the same. Thus, up to
this point, the renormalization of the noncommutative model preserves
supersymmetry.

\section{Renormalization of the $N$ point functions}

\begin{figure}
\centerline{\epsfbox{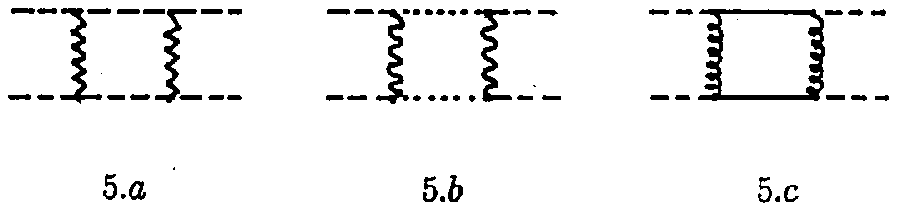}}
\caption{{Leading order contributions to the four point function of the
scalar field $\varphi$.}}
\label{Fig5}
\end{figure}

To complete our study of the renormalization program we focus next
on the four point function of the scalar field $\varphi$,
$\Gamma^{(4)}_{i_1i_2i_3i_4}(p_1,p_2,p_3,p_4)$, where the subscripts
are $O(N)$ indices.  Without loosing generality we choose
$i_1=i_3$ and $i_2=i_4$ but $i_1\not=i_2$. The  diagrams
are those in Fig. \ref{Fig5} whose associated amplitudes are
  
\begin{mathletters}\label{32}
\begin{eqnarray}
\Gamma^{(4)}_a &=& \int \frac{d^Dk}{(2\pi)^D} \frac{i}{(k+p_1)^2-M^2}\,
\frac{i}{(-k+p_2)^2-M^2}\,\, \Delta_\lambda(k)\,\Delta_\lambda(k+p_1-p_3)\,\,{\cal C}(k),\\ 
\Gamma^{(4)}_b &=& \int \frac{d^Dk}{(2\pi)^D} i^2\,\, \Delta_\sigma(k)\,\Delta_\sigma(k+p_1-p_3)\,{\cal C}(k),\\ 
\Gamma^{(4)}_c &=& -\int \frac{d^Dk}{(2\pi)^D}\,\, Tr\left [ \frac{i}{(\not \!\! k+
\not \!\!p_1)-M}\,S_\xi(k+p_1-p_3)\,
\frac{i}{(\not \!\! k-\not \!\!p_2)-M}\, S_\xi(k)\right ]\,{\cal C}(k)\,, 
\end{eqnarray}
\end{mathletters}

\noindent
where 

\begin{equation}
{\cal C}(k)\equiv \cos(k\wedge p_1)\cos(k\wedge p_2)\cos[(k+p_1)\wedge p_3]
\cos[(k-p_2)\wedge p_4] .\label{33}
\end{equation}

\noindent
In $D=3$ the above integrals are linearly divergent. However, as far as
the sum of these amplitudes is concerned, it is readily
seen that the leading divergence cancels and that the subleading one vanishes
under symmetric integration. These considerations hold irrespective
of the  presence of the cosine factors in the function ${\cal C}(k)$.
Thus, there is no UV/IR mixing that would eventually give rise to
infrared divergences. 

For $D=2$ each integral in (\ref{32}) is quadratically divergent and
in their sum a logarithmic divergence still survives. An explicit
calculation shows that this divergence cannot be removed by a
counterterm whose nonlocal structure arises from a combination of
Moyal products of $\varphi$ fields. Because of this, in the remaining of this section, we restrict our discussion to $D=3$. Further remarks on $D=2$ will be postponed to the 
conclusions.

\begin{figure}
\centerline{\epsfbox{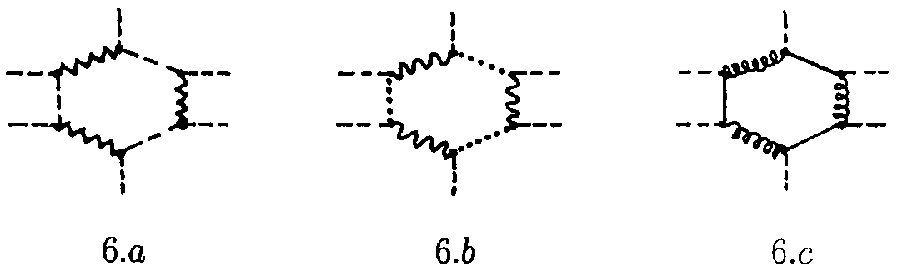}}
\caption{{Leading order contributions to the six point function of the
scalar field $\varphi$.}}
\label{Fig6}
\end{figure}

The graphs contributing to the six point vertex functions of the
$\varphi$ field are displayed in Fig. \ref{Fig6}. Since each graph in
the set has the same number of vertices, the cosine factor is the same
for all of them and, therefore, the cancellation among the divergent
parts proceeds as in the commutative case.

\begin{figure}
\centerline{\epsfbox{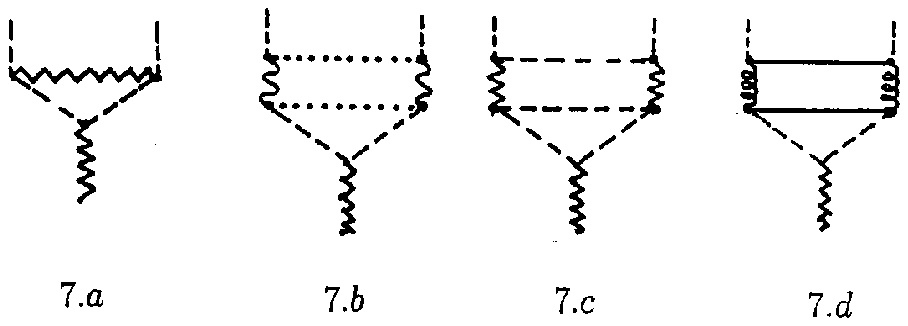}}
\caption{{Leading order contributions to the three point function 
$\Gamma_{\lambda\varphi\varphi}$. }}
\label{Fig7}
\end{figure}

As for the three point function, $\Gamma_{\lambda\varphi\varphi}$,
depicted in Fig. \ref{Fig7}, we notice that the diagrams can be grouped into
two sets according to the number of cosine factors. Indeed, diagram
7$a$ exhibits three cosine factors while diagrams 7$b$, 7$c$ and 7$d$
each contains five cosine factors. We have

\begin{equation}
\Gamma_{a}^{(3)}(p_1,p_2)=\int \frac{d^3k}{(2\pi)^3} \, I_a(k,p_1,p_2)\, {\cal C}_3 
\end{equation}

\noindent
and

\begin{equation}
\Gamma_{b+c+d}^{(3)}(p_1,p_2)=\int \frac{d^3k}{(2\pi)^3} \,\frac{d^3q}{(2\pi)^3} \, I_{b+c+d}(k,q, p_1,p_2)\, {\cal C}_5 
\end{equation}

\noindent
where $I_a$ and $I_{b+c+d}$ are the same integrands as in the
commutative case, and ${\cal C}_3$ and ${\cal C}_5$ are the cosine
factors

\begin{eqnarray}
{\cal C}_3 & =& \cos(k\wedge p_1) \cos(k\wedge p_2) \cos(k\wedge p_1 +
k\wedge p_2 + p_1\wedge p_2)
=\frac14 \cos (p_1\wedge p_2)+\ldots\\
 {\cal C}_5 & =& \cos(q\wedge p_1) \cos(q\wedge p_2) \cos(k\wedge p_1 +
k\wedge p_2 + p_1\wedge p_2) \cos(q\wedge p_1-k\wedge p_1+ q\wedge k)
\nonumber\\
&\phantom =&
\cos (k\wedge p_2 - q\wedge p_2 + q \wedge k) =
\frac1{16} \cos (p_1\wedge p_2)+\ldots,
\end{eqnarray}

\noindent
where the ellipsis indicate terms which give finite nonplanar contributions 
to the corresponding Feynman integrals.
Hence, the divergent part of each
set of graphs becomes altered with respect to the commutative regime
\cite{Koures} in such a way that it can no longer be absorbed by just
a wave function renormalization of the $\varphi$ field. A wave function
renormalization of the auxiliary field $\lambda$ is now needed. The
same applies to the three point functions $\Gamma_{\sigma \overline \psi\psi}$
and $\Gamma_{\overline \psi \xi \varphi}$. Supersymmetry requires that the
renormalization of $\sigma$, $\xi$ and $\lambda$ be the same, which can be
verified to be case. This is to be contrasted with the situation
in the commutative setting where no renormalization
is required for the Lagrange multiplier fields.

\section{Final remarks and conclusions}

In the commutative GN model the coupling constant plays a dual role,
eliminating divergences in both, the gap equation and the proper two point
function of the Lagrange multiplier field $\sigma$. Due to the presence
of the cosine factors, this property is no longer valid at the
noncommutative level and the $O(N)$ model becomes nonrenormalizable. The
noncommutative nonlinear sigma model is afflicted by a different type
of inconsistency, namely the UV/IR mixing which destroys its $1/N$
expansion. However, supersymmetry blends these models together and, for
$D=3$, leaves us with a consistent noncommutative quantum field
theory. 
It is worth mentioning that the noncommutative $U(N)$ GN model by
itself does not suffer from the above inconsistencies in the leading
order of $1/N$. Nevertheless, this does not apply to higher orders.

In $D=2$ the noncommutative supersymmetric nolinear  model possesses certain
peculiarities which are not present in $D=3$. Since the scalar field in
$D=2$ is dimensionless, any graph involving an arbitrary number of
external $\varphi$ lines is quadratically divergent. As exemplified by
the four point function $\Gamma^{(4)}_{i_1i_2i_3i_4}(p_1,p_2,p_3,p_4)$
of the $\varphi$ field, supersymmetry provides a partial cancellation
of divergences but a logarithmic one remains. The counterterm needed to
remove such divergence would have the form

\begin{equation}
 \int
\prod_{i=1}^4
\frac{d^2k_i}{(2\pi)^2}\cos[k_1\wedge k_2-k_3\wedge k_4]
(2\pi)^2\delta(k_1+k_2+k_3+k_4)\tilde \varphi_i(k_1)\tilde 
\varphi_j(k_2)\tilde 
\varphi_i(k_3)\tilde\varphi_j(k_4),\label{34}
\end{equation}

\noindent
where $\tilde \varphi_l(k)$ is the Fourier transform of $\varphi(x)$.
However, (\ref{34}) can not be written in terms of the two independent Moyal orderings 
$\int d^2 x \varphi_i\star\varphi_i\star\varphi_j\star\varphi_j$ and 
$\int d^2 x\varphi_i\star\varphi_j\star\varphi_i\star\varphi_j$.
Alternatively, one may entertain the possibility of canceling the
divergences by generalizing the definition of 1PI diagram as suggested
in \cite{Arefeva1} for the commutative nonlinear sigma model. Nevertheless,
the unbalance of cosine factors precludes this mechanism.
These observations cast doubts on the renormalizability of the noncommutative
supersymmetric $O(N)$ nonlinear sigma model in two space--time dimensions.

As remarked before, the formulation of the noncommutative
four--fermion interaction allows for a term of the form $\overline
\psi_i\star\overline\psi_j\star\psi_i\star\psi_j$. This is an
interesting problem since the $1/N$ expansion involves now an
auxiliary tensor field. In this connection, the $1/N$ expansion 
for matrix models in comutative
space has been discussed in the literature \cite{Hooft}

Summarizing, in this paper and in a previous one we have verified that
supersymmetrization corrects some of the difficulties arising
in the formulation of a noncommutative model as, for example the UV/IR
mixing. We would like to mention that another possibility for solving
the UV/IR problem, based on a resummation of the perturbative series,
has been considered in the literature \cite{Griguolo}.

\section{Acknowledgments}
This work was partially supported by Funda\c c\~ao de Amparo \`a
Pesquisa do Estado de S\~ao Paulo (FAPESP) and Conselho Nacional de
Desenvolvimento Cient\'\i fico e Tecnol\'ogico (CNPq). H.O.G and
V.O.R also acknowledge support from PRONEX under contract CNPq
66.2002/1998-99.

\input{epsf.tex}
\vspace*{3.0cm}
\vspace*{3.0cm}
\newpage
\vspace*{2.0cm}
\vspace*{2.0cm}
\vspace*{2.0cm}

\end{document}